\documentclass{sig-alternate}
\usepackage{verbatim}
\usepackage{epstopdf}

\newtheorem{definition}{Definition}

\newtheorem{theorem}{Theorem}

\begin{document}

%
\conferenceinfo{WOD}{2012 Nantes, France }

\title{Non-Interactive Differential Privacy: a Survey
\titlenote{An extended version of this paper will appear in the ACM International Conference Proceedings Series - ISBN number 978-1-4503-1404-6}}

%
%
%
%
%

\numberofauthors{1} 
%
\author{
%
%
\alignauthor
		David Leoni\\
		 \affaddr{LINA lab, University of Nantes}\\
       \email{david.leoni@etu.univ-nantes.fr}
\alignauthor
		David Leoni\\
		 \affaddr{LINA lab, University of Nantes}\\
       \email{david.leoni@etu.univ-nantes.fr}
}

\date{06 March 2012}

\maketitle
\begin{abstract}
OpenData movement around the globe is demanding more access to information 
which lies locked in public or private servers. As recently reported by a McKinsey
publication, this data has significant economic value, yet its release has potential
 to blatantly conflict with people privacy. 
Recent UK government inquires have shown concern from various parties about publication of 
anonymized databases, as there is concrete possibility of user
identification by means of linkage attacks. Differential privacy stands out
as a model that provides strong formal guarantees about the anonymity of the
participants in a sanitized database. Only recent results demonstrated its applicability on 
real-life datasets, though. This paper covers such breakthrough discoveries, 
by reviewing applications of differential privacy for non-interactive publication of anonymized real-life datasets. Theory, utility and a data-aware comparison are discussed on a variety of principles and concrete applications. 
\end{abstract}




\section{Introduction}
\nocite{}
\subsection{Motivation}
In a recent report by McKinsey \cite{McKinseyCons} it is estimated that in the developed economies 
of Europe alone, government administration could save more than 
 100 billion euros (149 billion dollars) in operational efficiency improvements alone by leveraging big data. This term refers to the enormous quantity of information organizations around the globe
collect daily. In particular, public institutions retain data about
many aspects of our life, including medical, fiscal, transportation and
criminal records. Private companies are
also increasingly taking a bigger role in our private life by recording our
Internet searches, friends network, financial transactions and transportation
habits. Not everybody knows how to handle this information properly, though. UK, the leading European country in terms of Open Data, recently held 
a consultation \cite{UKCons} with public institutions and industry representatives to discuss data publishing issues. Various parties expressed a clear concern about privacy  issues, prompted in part by clamorous episodes of privacy breaches occurred in the past. In 1997 Latanya Sweeney, proved 87\% of American citizens can be uniquely identified just by knowing their gender, ZIP code and
birthdate. To make a point of the claim she obtained this data from US public
voting records of Massachussets and linked it with supposedly ``anonymized''
hospital records of public employees. Next, she sent to the Governor of
Massachussets his medical records. The quest for true anonymization 
in the field of Privacy Preserving Data Publishing (PPDP) began,
and it is still not over. In 2006 Internet provider AOL released its search log containing 3 months of searches of 650,000 users. Usernames were masked with random identifiers, still, in a matter of days, a New York Times reporter identified Thelma Arnold, a
62-year old widow from Lilburn, GA as user \#4417749 {\cite{AOLNY}}, and her
queries became known to the world. As a consequence of releasing this private dataset the
CTO of AOL resigned, two employees were fired and a class action lawsuit is
pending. Later the same year, Netflix, a DVD rental company released a perturbed version of one tenth of its database of movie ratings expressed by its customers. A prize of 1,000,000\$ was offered to whoever improved by 10\% the accuracy of the company's own recommandation algorithm. The following year the researchers Narayanan and Shmatikov proved it was possible to identify users by linking them to Imdb, a public database of movie ratings in which users voluntarily can publish their ratings{\cite{Narayanan:Netflix}}. This concerns prevented in 2010 NetFlix from
proposing a follow-up of the prize.
\newpage
\subsection{Solutions}
Analysts want to have precise answers to queries about data, which can be sensitive. In the so-called \textit{interactive} setting, information is protected inside a database handled by the data owner, and access to it is allowed only through an interface. Answers provided by the interface are processed in such a way to guarantee the anonymity of the participants in the database. 
 There are two main problems with this approach. Suppose we have a database about HIV positive people containing their \textit{gender}, \textit{ZIP code} and \textit{birthdate}, along with a numerical \textit{id} and many other attributes. We might consider the identity of a patient at risk if anyone in the world gets to know at the same time at least three of his attributes. Then we might allow analyst $A$ to know i.e. \textit{gender} and \textit{ZIP} of person \#1, and analyst $B$ the \textit{ZIP} and the \textit{birthdate} of the same person. The system can answer both questions and privacy in this model is not in danger \textit{only as long as they don't share information}. If we wish to protect the data against collusion of data consumers as soon as two attributes about a given person are revealed to anybody we must disallow queries for all the remaining attributes. This can soon prevent the system to answer any query to new analysts. For example, if we allow each analyst to know at most one attribute per person, since future queries are unknown it might happen some analyst will never take advantage of his right to know one attribute for a given tuple thus wasting information others might be interested in. In the  \textit{non-interactive} setting these problems are addressed by releasing once and for all the data which we think is of interest to most analysts, while still preserving privacy. Naturally the example we made is simplistic and with this paper we intend to prove a wealth of useful information can be published while formally maintaining strong privacy guarantees.
Over the years, several solutions to solve the problem of protecting privacy
in anonymized databases have been proposed. Examples are
$k$-anonymity {\cite{Sweeney:Kanon}}, $l$-diversity {\cite{Mach:ldiv}},
$t$-closeness {\cite{Li:t-closeness}}. All these methods suppose it is worth to
distinguish data attributes into these groups: identifiers (i.e. name, surname),
quasi-identifiers(i.e. ZIP code, gender, age) and sensisitive (i.e. pathology,
rentedAdultMovie). In legal terms, in the EU the Data Protection Directives  \cite{EUPC95} define personal data as `information related to an identified or identifiable natural person'. It is a quite general definition, and for example even a house value can be classified as personal information as it might reveal its owner income. 
 Recently, the European Data Protection Supervisor EDPS  expressed its concerns \cite{EDPSRec} about a proposal on re-use of Public Sector Information (PSI) previously adopted by the  European Commission \cite{EC11}. In particular it was recommended that
 \begin{quote}
``Where appropriate, the data should be fully or partially anonymised and license conditions should specifically prohibit re-identification of individuals and re-use of personal data for purposes that may individually affect the data subjects.''
\end{quote}
The \textit{purpose limitation} is a difficult issue to solve in a context where PSI is put on the Internet for everybody to see. European transgressors who try to identify persons whose data is contained in a published anonymized dataset may be fined, but how to deal with non-European ones? Also, how is it possible to measure the degree of anonymization of a given dataset in order to decide if it is too risky to be published on the Internet? For example, the UK Office for National Statistics is going to release data collected in 2011  anonymized with a record-swapping system  \cite{UK_ONS_Eval}, which involves selecting households which are deemed too identifiable and swapping them with other households which are not too far in the same geographical region and have similar values.
 Tables containing origin-destination data are considered too hard to anonymize in a satisfying way so they are licensed only to restricted users. What are the theoretical basis for this distinction, if any?
The EDPS calls for a `proactive approach' which should be taken by authorities, meaning privacy issues should be analyzed at the earliest stages and involved people informed throughout all the data process release. 
 Linkage attacks shown before demonstrate how quasi-identifiers
can be used to significantly increase the accuracy in identity disclosure,
making the distinction with identifiers purely artificial. Also, sometimes the
sole fact of knowing somebody is or is not in a database may provide a malicious user with valuable information to carry out an attack.  So, how do we reach the so called \textit{privacy by design}, when a data release process is devised to prevent disclosure with formal guarantees? 
To respond to these issues the concept of \textit{differential privacy} was
introduced by Dwork {\cite{Dwork11}} to prevent attackers from being capable
even to detect the presence or absence of a given person in a database.
Differential privacy falls in the category of so called perturbative methods,
which attempts to create uncertainty in the released data by adding some
random noise. If database participants are independent from each other,
differential privacy promises that even if an attacker knows everything about
every user in the db but one, by looking at the published statistics he won't
be able to determine the identity of the remaining individual. Kieron O'Hara, in his 2011 independent transparency and privacy review to UK governement \cite{UKTranspGov}  mentions differential privacy as a cutting-edge technology that judges the \textit{computation} of the anonymization algorithm as privacy-preserving or otherwise, rather than trying to make an impossible distinction between identifying and non-identifying \textit{data}. This might sound promising, but O'Hara claims differential privacy  appears to be limited to the interactive setting. Is this really true? Recent results in the non-interactive setting are encouraging. In what follows, we formalize some concepts about differential privacy.

\subsection{Basic definitions}
We use $P(A)$ to indicate the probability of the occurrence of event $A$ and define  $\left\| x \right\|_1$ as the sum of all elements in vector $x$.

\begin{definition}[database] Given a database universe $\mathcal{D}$ we define a database $D \in \mathcal{D}$ as multiset of $\left| D \right|$ tuples from a universe $\mathcal{U}$ each with $k$ attributes. We say two databases  $D_1, D_2$ are neighbors if they differ in one tuple. We indicate such condition as  $\left| D_1 \Delta D_2 \right| = 1$
 
\end{definition}

\section{Differential privacy}
Randomized algorithms to publish sensitive data are called mechanisms. Since we are addressing the problem of statistical disclosure at
large, we use $\mathcal{R}$ to denote a wide range of output possibilities for
the mechanism designers, whose goal is to devise a mechanism function  $\mathcal{D} \rightarrow \mathcal{R}$. One possible choice of $\mathcal{R}$ could be
$\mathcal{D}$ itself, meaning we are going either to release a new database
composed by synthetic individuals who hopefully follow the same distribution
of the original participants or we publish a perturbed version of the original
database, with real data randomly modified to satisfy differential privacy
criteria. An another possible and popular choice of $\mathcal{R}$ is
the set of queries $q_j$ counting how many individuals $u_i$ satisfy a given
property $\gamma_j \left( u_i \right)$.
 A mechanism in order to be $\epsilon$-differentially private must satisfy the
following definition first introduced by Dwork {\cite{DwCalib}}, which in recent
years has become popular among researchers in the field of statistical
disclosure:
\begin{definition}[$\epsilon$-dp] Given a randomized mechanism $M : \mathcal{D}
  \rightarrow \mathcal{R}$ and a real value $ \epsilon > 0 $,  we say $M$ satisfies \\ $\epsilon$-differential
  privacy if $\forall D_1, D_2 \in \mathcal{D}$ such that $\left| D_1 \Delta
  D_2 \right| = 1$ and $\forall R \subseteq \mathcal{R}$ the following
  equation holds:
  \[ P \left( M \left( D_1 \right) \in R \right) \leqslant e^{\epsilon} P
     \left( M \left( D_2^{} \right) \in R \right) \]
\end{definition}

Differential privacy guarantees the following: a data release mechanism is
$\epsilon$-differentially private if, for any input database, any participant
$u$ in the database, and any possible output of the release mechanism $r$, the
presence or absence of participant $u$ (in db terms, $D_1 \ensuremath{\operatorname{and}} D_2$
differing for one row) causes at most a multiplicative $e^{\epsilon}$ change
in the probability of the mechanism outputting $r$. For example, if we want to release the count of people with HIV from a hypothetical medical database, we must devise a mechanism $\mathcal{L}$ that when executed on databases differing in one person probably outputs the same result. We can build such a mechanism by first counting the persons with a counting function $c:\mathcal{D} \rightarrow \mathbb{N} $ and then adding some noise to it. If the noise follows the Laplace distribution {\cite{DwCalib}}   
 we can have good outputs close to the true count at a rate exponentially greater than values far from it (see Fig. 1). To determine the amount of noise to add we must first introduce the concept of global sensitivity:
\begin{definition}[global sensitivity of a function]   We define the global
  sensitivity $\Delta \left( f \right)$ of a function $f : \mathcal{D}
  \rightarrow \mathbb{R}^w$, $w\in \mathbb{N^+}$,  as
  \[ \Delta \left( f \right) = \underset{\begin{array}{c}
       D_1, D_2 \in \mathcal{D}\\
       \left| D_1 \Delta D_2 \right| = 1
     \end{array}}{\max} \left\| f \left( D_1 \right) - f \left( D_2 \right)
     \right\|_1 \]
\end{definition}
A function has low sensitivity if the addition or removal of one person to the database can only change the outcome of the function evaluation by a small amount.  The so-called Laplace mechanism $\mathcal{L}$ works in fact for any numerical function $f: \mathcal{D} \rightarrow \mathbb{R}^w$ we want to compute on our database, but there is a catch: the amount of noise we must add is linked to the  global sensitivity of $f$. If we apply first $f$ on a db $D_1$, and then on a neighboring db $D_2$, if $f$ changes a lot it means we will need to add more noise to probably obtain the same output. For the single counting function $c$ the global sensitivity is low ($\Delta(c) = 1$) and thus the noise to add is limited.
  
\subsection{Differential privacy weaknessess}

\subsubsection{Relaxations}
Noise introduced by the randomization can produce results far from the true ones, thus leading to scarce utility of the published output for data consumers. Many relaxations of differential privacy exists to address this problem and the major one is $\left( \epsilon, \delta \right)$-differential privacy:

\begin{definition}[$\left( \epsilon, \delta \right)$-dp \cite{DOurData}] Given a randomized mechanism $M : \mathcal{D}
  \rightarrow \mathcal{R}$ we say $M$ satisfies $\left( \epsilon, \delta \right)$-differential
  privacy if $\forall D_1, D_2 \in \mathcal{D}$ such that $\left| D_1 \Delta
  D_2 \right| = 1$ and $R \subseteq \mathcal{R}$ the following equation holds:
  \[ P \left( M \left( D_1 \right) \in R \right) \leqslant e^{\epsilon} P
     \left( M \left( D_2^{} \right) \in R \right) + \delta \]
\end{definition}
There are no hard and fast rules for setting $\epsilon$ and $\delta$. It is generally left to the data releaser, and usually $\delta$ is taken to be very small, $\delta \leqslant 10^{- 4}$.  \ $\left( \epsilon, 0 \right)$-dp is the same as $\epsilon$-dp.
Among the other relaxations we mention $\left( \epsilon, \delta \right)$-probabilistic differential privacy ($\left( \epsilon, \delta \right)$-pdp) \cite{MKiMap}. A mechanism satisfying $\left( \epsilon, \delta \right)$-pdp satisfies also $\left( \epsilon, \delta \right)$-dp, but the converse does not hold. 

\subsubsection{Is differential privacy good enough?}

Some people say even differential privacy is not enough to adequately protect
individuals from data disclosure. Kifer and Machanavajjhala in {\cite{KiferLunch}} 
point out that differential privacy really works only if
individuals are truly independent from each other. When there is no
independence the participation of somebody in the db can be inferred just by
looking at other (supposedly known and in relation with the ``victim'')
entries.  As a consequence, they claim we are forced to take into
consideration adversarial knowledge, even if differential privacy apparently
freed us from such a burden. From a practical point of view, Dankar and El Emam \cite{Danker12} address several issues of differential privacy in the context of health care. They evidence a lack of real-life deployments of differentially private datasets, which might cause difficulties in assessing responsabilities if privacy breaches occur (was the $\epsilon$ value appropriate, who else used with success such an $\epsilon$? etc...). It might also be difficult to explain the level of anonymization guaranteed to patients, as $\epsilon$ is a parameter of a formula quite theoretical in nature. Furthermore, since published data is obtained through randomization, sometimes it may look hard to believe - i.e. a randomized census dataset may indicate there are people living at the center of a lake. As a consequence, analysts might be lead to mistrust the approach (or who applied it). 
\subsection{Mechanisms}

The two main mechanisms are the already described Laplace {\cite{DwCalib}} and the Exponential mechanism {\cite{SExpMech}}. The former is used when the output is numerical while the latter when outputs are not real or make no sense after adding noise. Other mechanisms are Li \textit{et al}'s matrix mechanism {\cite{LiMatrix}}, the geometric mechanism (a discretized version of the Laplace mechanism) by Ghosh \textit{et al} {\cite{GGeomMech}} and the Gaussian mechanism {\cite{DOurData}}.

\section{Measuring utility}

Broadly speaking, the utility of a mechanism is its capability to minimize the
error, which is a measure of the distance between original input
db/statistics on it and noisy output db/statistics . 
Only utility of restricted classes of queries can be guaranteed {\cite{Blu11}} in the
non-interactive setting.  Blum, Ligett, and Roth {\cite{Blu11}} showed that
in such setting it is possible to answer exponentially sized
families of counting queries so in this paper we will mostly look at solutions
for publishing data that are useful for such queries. However, the choice of
suitable statistics is a difficult problem as these statistics need to mirror
the sufficient statistics of applications that will use the sanitized
database, and for some applications the sufficient statistics are hard to
characterize. Popular approaches to measure utility are $\left( \alpha,
\beta \right)$-usefullness {\cite{Blu11}}, relative error with correction for
small queries {\cite{Xiao11,ChenTraj}} and without correction
{\cite{CormSpat, XiaoXY10}}, absolute error
{\cite{CormSparse,DingCubes,LiCompr}}, variance of the
error{\cite{CormSpat,Xiao11,DingCubes}}, euclidean distance {\cite{LiCompr,HMGeom}}.
In the following, we are going to define them more precisely.

\begin{definition}[$\left( \alpha, \beta \right) -$usefulness\cite{Blu11}] A
  privacy mechanism M is $\left( \alpha, \beta \right)$-useful for queries in
  class C if with probability $1 - \beta$, for every $Q \in C$ and every
  dataset $D \in \mathcal{D}$, for $\tilde{D}$=$M\left( D \right), \left| Q
  \left( \tilde{D} \right) - Q \left( D \right) \right| \leqslant \alpha$
\end{definition}

It is adopted in {\cite{Blu11},\cite{XiaoXY10} (only
for a basic cell based algorithm), and {\cite{BRSparse}}. $\left(
\alpha, \delta \right)$-usefulness is effective to give an overall estimation
of utility, but according to {\cite{Chen11}} fails to provide intuitive
experimental results. {\cite{Chen11,ChenTraj,Xiao11}}
experimentally measure the utility of sanitized data for counting queries by
relative error adopting this formula:

\begin{definition}[relative error]
  Let $Q$ be a query and $M : \mathcal{U} \rightarrow R$ a privacy mechanism. 
  We denote relative error as $\ensuremath{\operatorname{rel}} \left( Q \right) = \frac{\left| Q
  \left( \tilde{D} \right) - Q \left( D \right) \right|}{\max \left( Q \left(
  D \right), s \right)}$ where $s$ is a sanity bound that mitigates the
  effects of the queries with excessively small selectivities. In both
  {\cite{Xiao11}} and {\cite{Chen11}} $s$ is set to $0.1$\% of $\left|
  \mathcal{D} \right|$ \ 
\end{definition}

When the database is considered as a vector of reals (so $k = 1
, A_1 =\mathcal{R}$) the euclidean distance can be used as utility.
Li \textit{et al} in {\cite{LiCompr}} measure the error as the euclidean distance
between original and noisy database $\ensuremath{\operatorname{Err}} \left( D \right) = \left\| D -
M \left( D \right) \right\|_2$, claiming their mechanism is capable in such a way
 to guarantee the utility for any class of queries. Hardt \textit{et al} {\cite{HMGeom}} measure the
euclidean distance between query responses.

\section{Methods}
Several methods have been proposed to address the issue of releasing
differentially private data. Broadly speaking, they can be divided in the
categories of histogram construction, sampling and filtering, partitioning, dimensionality reduction. The notation $\tilde{O}$ indicates complexity with hidden logarithmic factors.

\subsection{Computing histograms }
A histogram is a disjoint partition of the database points with the number
of points which fall into each partition. Publishing a noisy version of the
histogram is appealing because of its usefullness for counting queries. However, the quality of queries
executed on the histogram may be low. If a
query requires the sum of $n$ histogram points, since each of them has some
noise the total noise sums up $n$ times and can quickly become intolerable.  Another issue regards $|\mathcal{U}|$ cardinality. As pointed out by {\cite{CormSparse}},
any data with several attributes $A_i$ leads to huge contingency matrices of
size $\Pi_i \left| A_i \right|$. Among the works suffering from this problem we
find {\cite{DNRR09,DingCubes,Xiao11,XiaoXY10,HayBoost,LiMatrix}}. {\cite{Xiao11}} operates a tranform on the counts and adds noise in the
wavelet domain in time $O(|\mathcal{U}|+|D|)$, and similar techniques via post-processing with overlapping
information are suggested in {\cite{HayBoost}}. Li \textit{et al} {\cite{LiMatrix}}
generalizes last two approaches with the introduction of the matrix mechanism
that generates an optimal query strategy based on the query workload of linear
count queries. No efficient algorithm is provided, though. One possible solution to the histogram problem is to take advantage of
sparsity of data present in many databases. This condition occurs when the number of cells $\left| \mathcal{U}^+ \right|$ with positive count in the contingency table in the database at hand is much bigger than zero-valued entries. To prove
this fact Cormode \textit{et al} in {\cite{CormSparse}} define sparsity $\rho$ as
$\rho = \left| \mathcal{U}^+ \right| / \Pi_i \left| A_i \right|$ .

\nocite{OnTheMap}
\nocite{CENSUS}
\nocite{UCIAdult}

\begin{figure}

\centering

\epsfig{file=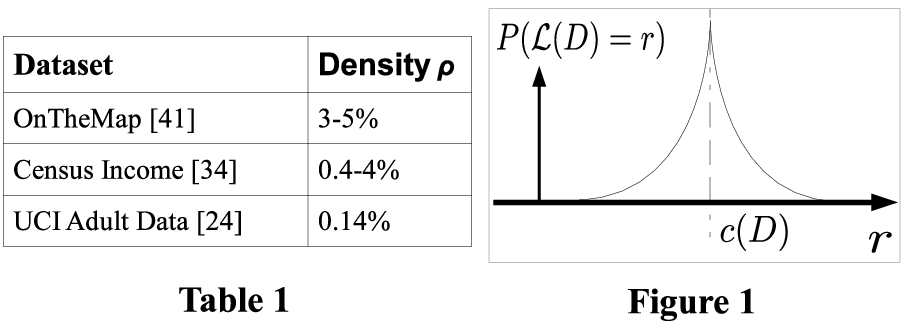,  width=8cm}
\label{fig:Laplace_example_1}
\end{figure}	

Table 1 is an example of the fact many natural datasets have low
density in the single-digit percentage range, or less. Applying differential
privacy naively generates output which is $1 / \rho$ times larger than the
data size. In the above examples, $1 / \rho$ ranges from tens to thousands.
which is clearly not practical for today's large data sizes. Among the methods which exploit data sparsity we
find  \cite{CormSparse,LiCompr,BRSparse}. In {\cite{BRSparse}} 
this definition of $m$-sparse queries is proposed:
\begin{definition}[$m$-sparse query\cite{BRSparse}] We say that a linear query $Q$
  is $m$-sparse if it takes non-zero values on only $m$ universe elements, and
  that a class of queries is $m$-sparse if each query it contains is
  $m'$-sparse for some $m' \leqslant m$. 
\end{definition}

\subsection{Sampling and filtering}
For the sampling and filtering category the idea is to avoid publishing huge
contingency tables by filtering out entries with small counts, which are often
in significant quantity in many databases. Cormode \textit{et al} {\cite{CormSparse}}
adopt a variety of filtering techniques - highpass filtering and priority
sampling being the most useful - to override the costly operation of
materializing a complete noise contingency table. Their method is suited for
sparse datasets.  For search log analysis in {\cite{KorSearch}} and
{\cite{GotzSL}} a mechanism is proposed to release noisy aggregated user query
and clicked url counts by filtering out excessively small counts. However,
such approaches break the association between distinct query-url pairs in the
output since all the user-IDs are removed, which might be useful in only a few
applications. Therefore, in {\cite{HongSL}} a sampling method is proposed to 
allow analysis in exactly the same fashion
and for the same purpose as the original data. However, $\left( \epsilon, \delta
\right)$-pdp is adopted to provide formal guarantess because relaxations are
indispensable in search log publishing as proven in {\cite{GotzSL}}.

\subsection{Partitioning}
Partitioning is indicated for ordered attributes such as spatial
data. Like in algorithms computing histograms, the universe $\mathcal{U}$ is divided into regions but in this case the shape of the cells is not fixed and an attempt is made to find an optimal subdivision of the space. Regions may be overlapping. The goal is to optimize the results of range queries, where the analyst asks for the number of people lying under a given query area, usually expressed as an hyperrectangle. This calculation involves the sum of already published noisy counts so a strategy to allow the user to minimize the total noise variance must also be provided.
A popular approach to partitioning is with $k d $-trees:
 at each round, an attribute is chosen and points in the database are split with some criteria. Usually uniformity in the number of points  on both sides of the splitting line is considered by choosing the median. Noisy counts of the two newly founded partitions are then published and partitioning is done recursively. 
 The idea of differentially private data-partitioning index
structures is suggested in the context of private record matching in {\cite{Inan2010}}. The approach there is based on using an approximate mean as a surrogate for median (on numerical data) to build $k d$-trees. The approach of {\cite{XiaoXY10}} imposes a fixed resolution grid over the base data. It then builds a $k d$-tree based on noisy counts in the grid, splitting nodes which are not considered `uniform', and then populates the final leaves with `fresh' noisy estimated counts. Quadtree partitioning simply imposes a recursive fixed grid in which at each round the space is divided into four rectangular cells of the same size. In {\cite{CormSpat}} a comparison between several median finding methods, Hilbert R-trees and quadtrees partitioning is performed and privacy budget is allocated in a geometrically increasing way to counts during the partitioning of 2D data. Attention is devoted in post-processing the noisy counts to achieve consistency and minimum error variance in time linear in the size of the published tree.  Quadtree partitioning is found to be fast and superior in quality of the output to all the other tested methods. 


\subsection{Dimensionality reduction}
Dimensionality reduction methods usually consider the database as a matrix and
apply random projections on it. In this line of research we find {\cite{BRSparse}} in which for the class of linear counting queries that are $m$-sparse 
a method based on releasing a perturbed random projection of the private
database together with the projection matrix is described. Running time
is polynomial in the database size $\left| D \right|$, $m$, and $\log \left| \mathcal{U} \right|$. In {\cite{ZhouCompr}} compression is applied to obtain a reduced syntethic database $D'$ of size $\left|
D' \right| \ll \left| D \right|$ in polynomial time. Li et
al {\cite{LiCompr}} apply compressive sensing to
obtain a perturbed database from sparse data through decompression in time
$\tilde{O} \left( \left| D \right| \right)$. 

\section{Applications}
In recent years differential privacy has been successfully applied to a wide range of real-world data, although generally with no quality assessment by final users of anonymized datasets. In {\cite{MKiMap}} $\left(
\epsilon, \delta \right)$-pdp is introduced to model spatial data. 
This solution is then compared by Cormode with his work in \cite{CormSparse}. In Y.Xiao \textit{et al} {\cite{XiaoXY10}} a $k d$-tree technique is applied on CENSUS data
{\cite{CENSUS}}, and results are found superior to Inan \textit{et al} hierarchical tree
method {\cite{Inan2010}}. Moreover, the open source HIDE platform {\cite{HIDE}} is
provided to experiment with four differentially
private algorithms: {\cite{HayBoost,Inan2010,Xiao11,Chen11}}. Cormode later in {\cite{CormSpat}} found his algorithm to
give less error than Inan's {\cite{Inan2010}} and Y. Xiao works
{\cite{XiaoXY10}}. In {\cite{Chen11}} MSNBC {\cite{MSNBC}} and STM {\cite{STM}} datasets represented as set-valued
boolean data are considered. The only comparison is performed for MSNBC against basic noisy datacube method of Dwork's{\cite{DwCalib}}, as STM has big universe $\left|
\mathcal{U} \right|$ size and few methods are capable to handle this
situation. STM dataset represented as sequences of locations is also considered in {\cite{ChenTraj}}, although location coordinates nor time intervals are taken into account. In \cite{GotzSL}  publication of counting queries for search logs is considered, but dataset origin is not specified. In \cite{HongSL} AOL search log \cite{AOLNY} is adopted for experimental tests. \cite{Xiao11} performs experiments on CENSUS data \cite{CENSUS} using binning to have $|\mathcal{U}|\approx $16,000,000.

\section{Synthetic databases}

There have been few attempts to devise mechanisms of the kind $M : \mathcal{D}
\rightarrow \mathcal{D}$, because privacy in these cases is more difficult to
preserve. Outputs can be either a synthetic database - in which individuals
follow the same distribution as in the original database - or just a perturbed
version, where rows are directly taken from the original database with some
modification to guarantee anonymity. Perturbed database release is considered in  {\cite{Chen11,ChenTraj,LiCompr}}. Synthetic data is released with methods proposed in  {\cite{ZhouCompr,MKiMap,HongSL}}.

\section{Conclusions}

Differential privacy provides formal guarantees
that public opinion needs when privacy is at stake, yet for many years such requirements
were judged by researchers too strict to be applicable. Recently, several breakthrough results changed this mood. We presented a variety of methods -
partitioning, dimensionality reduction, sampling and filtering - which have
been successfully applied to many real-life datasets. Some methods were also
shown for histogram publishing, which, albeit unfeasibile on certain datasets
with big universe size, can still be used in practice on some real life
datasets. Most of the papers we discussed about use a plain $\epsilon$-dp
model which seems to indicate relaxations may not really be needed except in
 problematic cases like search log publishing. Differential privacy can be applied efficiently 
with formal guarantees to set-valued data {\cite{Chen11}},  to sparse data for counting queries {\cite{BRSparse}} 
and for general purpose queries {\cite{LiCompr}}. When data is not
sparse and $|\mathcal{U}|$ is not too big \cite{Xiao11} can be used with success.
 For the difficult case of search log publishing Hong \textit{et al} \cite{HongSL} showed it is even possible to publish a perturbed database while maximizing utility. Less formal guarantees but
good practical results are provided efficiently in
{\cite{ChenTraj}} for sequences of short length and Cormode {\cite{CormSparse}} for discrete data.  
Good results were obtained for bidimensional spatial data in {\cite{CormSpat}}. 
For these reasons time is ripe for the Open Data movement
to start considering the adoption of differential privacy and provide people with
adequate guarantees about the way their data is handled. 
Research has still to be done to impose constraints on output data in order to avoid inconsistencies and to properly anonymize highly dimensional non-sparse data and
preserving utility of general classes of queries. In this regard, publication of synthetic or perturbed datasets seems a promising approach, which needs careful query utility examination.

\bibliographystyle{acm}

\begin{thebibliography}{}

\bibitem{AOLNY}
{\sc Barbaro, M., and Zeller, T.}
\newblock A face is exposed for aol searcher no. 4417749.
\newblock {\em New York Times\/} (2006).

\bibitem{Blu11}
{\sc Blum, A., Ligett, K., and Roth, A.}
\newblock A learning theory approach to non-interactive database privacy.
\newblock {\em CoRR abs/1109.2229\/} (2011).

\bibitem{BRSparse}
{\sc {Blum}, A., and {Roth}, A.}
\newblock Fast private data release algorithms for sparse queries.
\newblock {\em ArXiv e-prints\/} (nov 2011).

\bibitem{ChenTraj}
{\sc Chen, R., Fung, B. C.~M., and Desai, B.~C.}
\newblock Differentially private trajectory data publication.
\newblock {\em CoRR\/} (2011), --1--1.

\bibitem{Chen11}
{\sc Chen, R., Mohammed, N., Fung, B. C.~M., Desai, B.~C., and Xiong, L.}
\newblock Publishing set-valued data via differential privacy.
\newblock {\em PVLDB 4}, 11 (2011), 1087--1098.

\bibitem{CormSparse}
{\sc Cormode, G., Procopiuc, C.~M., Srivastava, D., and Tran, T. T.~L.}
\newblock Differentially private publication of sparse data.
\newblock {\em CoRR abs/1103.0825\/} (2011).

\bibitem{CormSpat}
{\sc Cormode, G., Procopiuc, M., Shen, E., Srivastava, D., and Yu, T.}
\newblock Differentially private spatial decompositions.
\newblock {\em CoRR abs/1103.5170\/} (2011).

\bibitem{Danker12}
{\sc Dankar, F.~K., and El~Emam, K.}
\newblock The application of differential privacy to health data.
\newblock In {\em PAIS '12}.

\bibitem{DingCubes}
{\sc Ding, B., Winslett, M., Han, J., and Li, Z.}
\newblock Differentially private data cubes: optimizing noise sources and
  consistency.
\newblock In {\em Proc. of SIGMOD '11}, ACM, pp.~217--228.

\bibitem{Dwork11}
{\sc Dwork, C.}
\newblock A firm foundation for private data analysis.
\newblock {\em Commun. ACM 54}, 1 (2011), 86--95.

\bibitem{DOurData}
{\sc Dwork, C., Kenthapadi, K., {McSherry}, F., Mironov, I., and Naor, M.}
\newblock Our data, ourselves: Privacy via distributed noise generation.
\newblock In {\em EUROCRYPT '06}, LNCS, Springer, pp.~486--503.

\bibitem{DwCalib}
{\sc Dwork, C., McSherry, F., Nissim, K., and Smith, A.}
\newblock Calibrating noise to sensitivity in private data analysis.
\newblock In {\em TCC'06\/} (2006), pp.~265--284.

\bibitem{DNRR09}
{\sc Dwork, C., Naor, M., Reingold, O., Rothblum, G.~N., and Vadhan, S.}
\newblock On the complexity of differentially private data release: efficient
  algorithms and hardness results.
\newblock In {\em Proc. of STOC '09}, ACM.

\bibitem{EC11}
{\sc {European Commission}}.
\newblock Proposal for a directive of the european parliament and of the
  council {COM(2011)} 877 final, December 2011.

\bibitem{EDPSRec}
{\sc {European Data Protection Supervisor (EDPS)}}.
\newblock Opinion {EDPS/08/12}, Apr. 2012.

\bibitem{EUPC95}
{\sc {European Parliament}}.
\newblock Directive {95/46/EC} {(OJ L 281/95)}, October 95.

\bibitem{GGeomMech}
{\sc Ghosh, A., Roughgarden, T., and Sundararajan, M.}
\newblock Universally utility-maximizing privacy mechanisms.
\newblock In {\em Proc. of STOC '09}, ACM, pp.~351--360.

\bibitem{GotzSL}
{\sc Gotz, M., Machanavajjhala, A., Wang, G., Xiao, X., and Gehrke, J.}
\newblock Publishing search logs: A comparative study of privacy guarantees.
\newblock {\em IEEE TKDE. 24}, 3 (Mar. 2012), 520--532.

\bibitem{HMGeom}
{\sc Hardt, M., and Talwar, K.}
\newblock On the geometry of differential privacy.
\newblock In {\em Proc. of STOC '10}, ACM, pp.~705--714.

\bibitem{HayBoost}
{\sc Hay, M., Rastogi, V., Miklau, G., and Suciu, D.}
\newblock Boosting the accuracy of differentially private histograms through
  consistency.
\newblock {\em Proc. of VLDB Endow. 3\/} (September 2010), 1021--1032.

\bibitem{HongSL}
{\sc Hong, Y., Vaidya, J., Lu, H., and Wu, M.}
\newblock Differentially private search log sanitization with optimal output
  utility.
\newblock {\em CoRR abs/1108.0186\/} (2011).

\bibitem{Inan2010}
{\sc Inan, A., Kantarcioglu, M., Ghinita, G., and Bertino, E.}
\newblock Private record matching using differential privacy.
\newblock In {\em Proc. of EDBT '10}, ACM, pp.~123--134.

\bibitem{KiferLunch}
{\sc Kifer, D., and Machanavajjhala, A.}
\newblock No free lunch in data privacy.
\newblock In {\em Proc. of SIGMOD '11}, ACM, pp.~193--204.

\bibitem{UCIAdult}
{\sc Kohavi, R., and Becker, B.~D.}
\newblock http://archive.ics.uci.edu/ml/datasets/Adult.

\bibitem{KorSearch}
{\sc Korolova, A., Kenthapadi, K., Mishra, N., and Ntoulas, A.}
\newblock Releasing search queries and clicks privately.
\newblock In {\em Proc. of WWW '09}, ACM, pp.~171--180.

\bibitem{LiMatrix}
{\sc Li, C., Hay, M., Rastogi, V., Miklau, G., and McGregor, A.}
\newblock Optimizing linear counting queries under differential privacy.
\newblock In {\em Proc. of PODS '10\/} (2010), ACM, pp.~123--134.

\bibitem{Li:t-closeness}
{\sc Li, N., and Li, T.}
\newblock t-closeness: Privacy beyond k-anonymity and l-diversity.
\newblock In {\em In Proc. of IEEE ICDE ' 07}.

\bibitem{LiCompr}
{\sc Li, Y.~D., Zhang, Z., Winslett, M., and Yang, Y.}
\newblock Compressive mechanism: utilizing sparse representation in
  differential privacy.
\newblock In {\em Proc. of WPES '11}, ACM, pp.~177--182.

\bibitem{MKiMap}
{\sc Machanavajjhala, A., Kifer, D., Abowd, J.~M., Gehrke, J., and Vilhuber,
  L.}
\newblock Privacy: Theory meets practice on the map.
\newblock In {\em Proc. of ICDE'08}, IEEE, pp.~277--286.

\bibitem{Mach:ldiv}
{\sc Machanavajjhala, A., Kifer, D., Gehrke, J., and Venkitasubramaniam, M.}
\newblock L-diversity: Privacy beyond k-anonymity.
\newblock {\em ACM TKDD 1\/} (March 2007).

\bibitem{McKinseyCons}
{\sc {McKinsey Global Institute}}.
\newblock {Big data: The next frontier for innovation, competition, and
  productivity}, 2011.

\bibitem{SExpMech}
{\sc McSherry, F., and Talwar, K.}
\newblock Mechanism design via differential privacy.
\newblock In {\em Proc. of FOCS '07}, IEEE Computer Society, pp.~94--103.

\bibitem{MSNBC}
{\sc {Microsoft}}.
\newblock
  http://archive.ics.uci.edu/ml/datasets\\/MSNBC.com+Anonymous+Web+Data.

\bibitem{CENSUS}
{\sc {Minnesota Population Center (MPC)}}.
\newblock http://www.ipums.org.

\bibitem{Narayanan:Netflix}
{\sc Narayanan, A., and Shmatikov, V.}
\newblock Robust de-anonymization of large sparse datasets.
\newblock In {\em Proc. of SP '08}, IEEE Computer Society, pp.~111--125.

\bibitem{UKTranspGov}
{\sc O'Hara, K.}
\newblock Transparent government, not transparent citizens: A report on privacy
  and transparency for the cabinet office, Sept. 2011.

\bibitem{STM}
{\sc {Societè de transport de Montrèal}}.
\newblock http://www.stm.info.

\bibitem{Sweeney:Kanon}
{\sc Sweeney, L.}
\newblock k-anonymity: a model for protecting privacy.
\newblock {\em Int. J. Uncertain. Fuzziness Knowl.-Based Syst. 10\/} (October
  2002), 557--570.

\bibitem{UKCons}
{\sc {UK Cabinet office}}.
\newblock Making open data real: A public consultation, 2011.

\bibitem{UK_ONS_Eval}
{\sc {UK Office for National Statistics}}.
\newblock Evaluating a statistical disclosure control (sdc) strategy for 2011
  census outputs, 2011.

\bibitem{OnTheMap}
{\sc {US Census Bureau}}.
\newblock US Census Bureau, http://lehdmap.did.census.gov/.

\bibitem{Xiao11}
{\sc Xiao, X., Wang, G., and Gehrke, J.}
\newblock Differential privacy via wavelet transforms.
\newblock {\em IEEE TKDE 23}, 8 (2011), 1200--1214.

\bibitem{XiaoXY10}
{\sc Xiao, Y., Xiong, L., and Yuan, C.}
\newblock Differentially private data release through multidimensional
  partitioning.
\newblock In {\em Proc. of SDM '10}, Springer-Verlag, pp.~150--168.

\bibitem{HIDE}
{\sc Xiong, L., and Gardner, J.}
\newblock http://www.mathcs.emory.edu/hide/index.html.

\bibitem{ZhouCompr}
{\sc {Zhou}, S., {Ligett}, K., and {Wasserman}, L.}
\newblock {Differential Privacy with Compression}.
\newblock {\em ArXiv e-prints\/} (Jan. 2009).

\end{thebibliography}
 
\end{document}